\documentclass[twocolumn,preprintnumbers,amsmath,amssymb]{revtex4}
\usepackage{graphicx}
\usepackage{dcolumn}
\usepackage{bm}
\begin{document}

\title {Hopping induced continuous diffusive 
dynamics below the non-ergodic transition
}
\author{ Sarika Maitra Bhattacharyya$^{\dagger}\footnote
{Electronic mail~:sarika@sscu.iisc.ernet.in}$, 
Biman Bagchi$^{\dagger}$\footnote{Electronic mail~:bbagchi@sscu.iisc.ernet.in} \\
and \\
 Peter G. Wolynes$^{\ddagger}$\footnote
{Electronic mail~:pwolynes@chem.ucsd.edu}}

\affiliation{$^{\dagger}$ Solid State and Structural Chemistry Unit, 
Indian Institute of Science, Bangalore 560 012, India.\\
$^{\ddagger}$Department of Chemistry and Biochemistry, 
University of California at San 
Diego, La Jolla, California 92093-0371}

\begin{abstract}

In low temperature supercooled liquid, below the ideal mode coupling theory transition temperature, hopping and continuous diffusion are seen to coexist.
We present a theory which incorporates interaction between the two processes and shows that hopping can induce continuous diffusion in the otherwise frozen liquid. Several universal features arise from nonlinear interactions between the
continuous diffusive dynamics (described here by the mode coupling theory (MCT)) and the activated hopping (described here by the random first order transition theory ).
We apply the theory to real systems (Salol) to show that 
the theory correctly predicts the temperature dependence 
of the non-exponential stretching parameter, $\beta$, and the primary $\alpha$ relaxation timescale, $\tau$.   
The study explains why, even below the ergodic to non-ergodic transition, the dynamics is well
described by MCT. The non-linear coupling between the two dynamical processes modifies the
relaxation behavior of the structural relaxation from what would be predicted by a theory with 
a complete static Gaussian barrier distribution in a manner that may be 
described as a facilitation effect. 
Furthermore, the theory explains 
the observed variation of the stretching exponent $\beta$
with the fragility parameter, $D$.
\end{abstract}
\maketitle
\section{Introduction}
The glass transition is characterized by a number of interesting kinetic phenomena. 
Slow non-exponential relaxation of time correlation functions over large time windows 
is one such important phenomenon. 
This relaxation is usually well approximated by the stretched 
exponential, Kohlrausch-William-Watts (KWW) formula \cite{KWW}, $\phi(t)=exp(-(t/\tau))^{\beta}$, with both $\beta$ and $\tau$ exhibiting non trivial temperature dependence. 
The origin of the stretching is usually attributed to the presence of dynamic heterogeneity in the system
\cite{arbe1to5,arbe6}. The temperature dependence of the relaxation time can be described by the the Vogel-Fulcher-Tamman (VFT) expression, $\tau=\tau_{VFT}e^{DT_{o}/(T-T_{o})}$, 
where $\tau_{VFT}$ is the high temperature relaxation time, $T_{o}$ is the VFT temperature 
and $D$ is the fragility index. The fragility index, $D$ determines
the degree of deviation from the Arrhenius law. 
Experimental and theoretical model studies have shown that $\beta$ and $D$ are
correlated
\cite{bohmer,xiawoly}. The temperature dependence of $\tau$ is also described by phenomenological MCT expression,
$\tau\sim\left[(T-T^{fit}_{c})/T^{fit}_{c}\right]^{-\gamma}$, ($\gamma >0$). $T^{fit}_{c}$ is referred to as the MCT
transition temperature. Above $T^{fit}_{c}$, MCT is found to explain many experimental results \cite{cu1,cu2,cummins1,richert,fayer}, and below $T^{fit}_{c}$, the 
MCT description of continuous diffusion breaks down. It is conjectured that this breakdown is due to the ergodic to non-ergodic transition in the dynamics and below $T^{fit}_{c}$ activated dynamics becomes a dominant mode
of transport. However in an elegant work, Burmer and Reichman \cite{reich} (BR) have recently shown that the 
idealized MCT breaks down at a much higher
temperature, $T^{0}_{c}$, which corresponds to the temperature, $T_{L}$, where the landscape properties change. Kob {\it et al.} have
shown that the structural MCT predicts the proper dynamics till very close to $T^{fit}_{c}$ but needs static inputs to be calculated at a higher effective
temperature \cite{kob}. 
 Wolynes and co-workers 
have shown that below the ergodic to non-ergodic transition but
before the predicted crossover to the activated dynamics, string or fractal excitations drive the dynamics \cite{stevwoly}. 

Computer simulation studies predict the \textit{coexistence} of continuous diffusion and hopping as mechanism
of mass transport at temperatures much above $T^{fit}_{c}$\cite{sarikajcp,denny}.
These studies show that individual hopping events are often followed by enhanced
continuous diffusion or more hops of the surrounding atoms or molecules \cite{sarikajcp,pinakiprl}. Simulations have also shown that a single hopping event relaxes the local stress \cite{sarikajcp}, hence it is expected that hopping events are followed by continuous diffusive dynamics. As mentioned earlier, from BR study we know the temperature where the ergodic to non-ergodic transition takes place ($T^{0}_{c}=T_{L}$) \cite{reich}.
However, there is no clear theoretical 
understanding of the dynamics in the range, $T^{fit}_{c}<T<T^{0}_{c}$. Furthermore, we need to understand, why below $T^{0}_{c}$ the 
MCT still seems to explain the dynamics rather well and finally, what happens at $T^{fit}_{c}$ that 
leads to the breakdown of the MCT.
 
In an earlier work it was shown that indeed below the ergodic to non-ergodic transition, both activated and continuous diffusion mechanisms are present. These two are nonlinearly coupled and may drive each other\cite{sbp}.
It was shown that the full dynamics is a synergetic effect of continuous diffusive motion and activated dynamics \cite{sbp}. 

Here we show that the above coupling can lead to quite interesting phenomena. 
First for the Salol system we find that, the theory predicts the correct temperature dependence of relaxation time. Following BR analysis \cite{reich} we consider that the ergodic to non-ergodic transition in the ideal MCT occurs at $T^{0}_{c}=278 k$.
However, when we fit the relaxation timescale to the phenomenological MCT 
expression, the timescale shows an apparent divergence at a lower temperature, thus predicting $T^{fit}_{c}=258K$ which is similar to that found in experimental studies\cite{cu1}. 
Thus the study demonstrates that well below the idealized MCT
ergodic to non-ergodic transition temperature where non-mean field activated events are present, the dynamics 
can be well described by MCT. We find that, across $T^{fit}_{c}$, although both the transport mechanisms are operative and there is no discontinuity in their motion, there is an abrupt or sudden 
rise in the contribution from the activated events. 
This leads to the apparent breakdown of the MCT. 

Second, the theory predicts the correct temperature dependence of the stretching parameter
  $\beta$. 
We find that the bare barrier height distribution that describes isolated hopping, gets renormalized due to the coupling of the hopping dynamics to the continuous diffusive motion.
The renormalized distribution of the barriers, referred to as the dynamic barrier height distribution, is different from the initial assumption of distribution 
referred to as the static barrier height distribution. There is a preference for low barrier hopping events leading to 
facilitation effect. 
Third, due to facilitation effect, the theory predicts a correct relationship between the stretching parameter, $\beta$, and the fragility index, $D$.  

The organization of the rest of the paper is as follows. In the next section we
describe the theoretical scheme. Section 3
contains results and discussions.
Section 4 concludes with a discussion on the results.

\section{Theoretical formulation}

The present work uses a scheme 
of calculation similar to the one presented earlier \cite{sbp} with a few modifications.
For describing the diffusive motion we use $F_{12}$ model of the MCT \cite{gotzebook} and we also describe
the activated hopping dynamics with a static barrier height distribution \cite{xiawoly}. 
In describing the dynamics the coupling between the different wave-vectors are neglected and 
the contribution from the static and dynamic quantities are calculated at a single wavenumber $q=q_{m}$ ($q_{m}$ is the wave number where the peak of the structure factor appears) which 
is known to provide the dominant contribution.  The MCT part of the intermediate scattering function, $\phi_{MCT}(t)$ is self consistently calculated 
with the full scattering function, $\phi(t)$. In writing the equation of motion we consider 
that there are distribution of
hopping barriers in the system arising from the entropy fluctuation \cite{xiawoly}. We know that in the 
absence of hopping motion the MCT dynamics below certain temperature is completely 
frozen, leading to  
a separation of time scale between the hopping dynamics and MCT dynamics. Thus for the frozen MCT dynamics, it appears that multiple relaxation channels (present due to different hopping barriers) open up  and all these channels act in parallel to relax the system. Thus when the contribution from these multiple hoppings are included in the total relaxation, the equation of 
motion of MCT part of the intermediate structure function, $\phi_{MCT}(t)$, can be written as,
\begin{eqnarray}
&&\ddot{\bf\phi}_{MCT}(t) 
+\gamma \dot{\bf \phi}_{MCT}(t) 
+ \Omega_{0}^{2}
{\bf\phi}_{MCT}(t)\nonumber \\
&+& \lambda_{1} \Omega_{0}^{2} 
\int_{0}^{t} \:dt^{\prime} {\bf\phi}_{MCT}(t^{\prime}) \sum \phi^{s}_{hop}(t^{\prime})
\dot{\bf\phi}_{MCT}(t-t^{\prime}) \nonumber \\
&+& \lambda_{2} \Omega_{0}^{2} 
\int_{0}^{t} \:dt^{\prime}[{\bf\phi}_{MCT}(t^{\prime}) \sum \phi^{s}_{hop}(t^{\prime})]^{2}  
\dot{\bf\phi}_{MCT}(t-t^{\prime})\nonumber \\ 
&=& 0 \label{fqtmctdist}
\end{eqnarray}
\noindent 
Eq.\ref{fqtmctdist} is essentially of the same general form as derived by Sjogren and Gotze
for the extended mode coupling theory \cite{gotze}. This point has been elaborately analyzed in an earlier work \cite{sbp}.
In deriving equation \ref{fqtmctdist} we consider that the total intermediate scattering function is written as, 
\begin{equation}
\phi(t)\simeq \phi_{MCT}(t) \sum \phi^{s}_{hop}(t). \label{fqtfull}
\end{equation} 
\noindent
Here $\phi^{s}_{hop}(t)=\frac{1}{s+K_{hop}(q,\Delta F)}$ describes the activated hopping dynamics for a single hopping barrier. $K_{hop}(q,\Delta F)= {\cal F}(q)P_{hop}(\Delta F)$. 
The expression for ${\cal F}(q)$ is given by, \cite{correc} ${\cal F}(q)=\frac{v_{0}}{v_{P}} (1- G(q))$.
For the present work  the $q$ dependence of ${\cal F}(q)$ will 
be neglected and ${\cal F}(q)$  will be set to unity. $P_{hop}(\Delta F)$ is the average hopping
rate which is a function of the free energy barrier height, $\Delta F$ 
and is given by $P_{hop}(\Delta F)=\frac {1}{\tau_{0}}exp(-\Delta F/k_{B}T)$ \cite{lubwoly}. 
The total contribution from multiple barrier hopping motion is written as, $\sum \phi^{s}_{hop}(t)
=\int \phi^{s}_{hop}(t) {\cal P}^{static}(\Delta F) d\Delta F =
\int e^{-tK_{hop}(\Delta F)} {\cal P}^{static}(\Delta F) d\Delta F$ where ${\cal P}^{static}(\Delta F)$ is considered to be Gaussian, ${\cal P}^{static}(\Delta F)=
\frac{1}{\sqrt 2{\pi\delta\Delta F^{2}}}e^{-(\Delta F-\Delta F_{o})/2\delta\Delta F^{2}}$.
We will call this distribution as the static barrier height distribution.
 
We now discuss the MCT part of the calculation.
As in our earlier model 
calculation \cite{sbp}, the values of $\Omega_{o}$ and $\gamma$ 
are kept fixed at unity, neglecting their temperature dependence and the scaling time is taken as 1ps.  $
\lambda_{1}= \frac {(2 \lambda -1)}{\lambda^{2}}+ \epsilon \frac {\lambda}
{(1+(1-\lambda)^{2})}$ and $\lambda_{2}= \frac {1}{\lambda^{2}}+ \epsilon \frac {\lambda (1-\lambda)} 
{(1+(1-\lambda)^{2})}$ \cite{gotze}. The MCT formalism predicts a relationship between 
$\lambda$  and $\beta_{MCT}$ as $\beta_{MCT}=-log(2)/log(1-\lambda)$. $\epsilon$ is a measure 
of the distance from the ergodic to non-ergodic transition temperature of the ideal MCT, $T^{0}_{C}$, thus $\epsilon=\frac{T^{0}_{c}-T}{T^{0}_{c}}$.
Thus to calculate the MCT part of the relaxation we need to estimate $\lambda$ and  $T^{0}_{c}$. These quantities can be calculated for systems where the static 
quantities (like static structure factor) are known, but for realistic system due to the unavailability of the static quantities the estimation becomes difficult. We thus use the following methods to estimate $\lambda$ and  $T^{0}_{c}$. 
According to the analysis of BR, the ergodic to non-ergodic transition takes place 
at the temperature, $T_{L}$ where the energy landscape properties first change \cite{reich}, thus $T^{0}_{c}=T_{L}$. 
According to Sastry {\it et al.}\cite{sastry}, $T_{L}$ also coincides with the temperature where the stretching parameter starts falling. From experimental studies we know that for Salol system the stretching parameter starts falling 
at $T=278K$ \cite{cu2}. Thus we estimate that $T^{0}_{c}=T_{L}=278K$. 
Now for the estimation of $\lambda$ we again make use of experimental results. MCT is expected to explain the dynamics above $T^{0}_{c}$, thus the stretching parameter above $T^{0}_{c}$ should be equal to $\beta_{MCT}=0.84$ \cite{cu2}.
We have also mentioned that $\lambda$ and $\beta_{MCT}$ are related. Thus $\lambda$ is fixed in such a way that, above $T^{0}_{c}$, we get the correct $\beta_{MCT}$.

Next we discuss the implementation of the hopping dynamics
for Salol system using the predictions of random first order transition (RFOT) theory.
According to the RFOT theory, the mean barrier, $F(r)$, which is used to calculate the mean barrier height, can be written as $F(r)= \frac{\Gamma_{K}(r) \Gamma_{A}(r)}{\Gamma_{K}(r)+ \Gamma_{A}(r)} -\frac {4 \pi}{3} r^{3} T s_{c}$ 
where $\Gamma_{K}(r)$ and $\Gamma_{A}(r)$ are the surface energy terms at 
$T_{K}$ (Kauzmann temperature) and $T_{A}$ (temperature where hopping barrier disappears) respectively \cite{sbp,lubwoly}.  
The temperature dependence of the configurational entropy can also be given by an empirical formula,\cite{angel} $s_{c}=s_{fit}(1-T_{K}/T)$  where $s_{fit}$ is a system dependent parameter which is also related to the 
specific heat jump at the Kauzmann temperature (${\bar \Delta}c_{p}(T)=s_{fit} (T_{K}/T)$). 
For the Salol system $s_{fit}=2.65, T_{K}=175 K$ and $T_{A}=333K$. With these values of the parameters 
the mean barrier height and the critical nucleus radius are calculated. We find that at $T=280K$, the size of 
the critical nucleus is above unity. Thus although $T_{A}=333K$, for all practical purpose $T=280K$ should be considered as the temperature where activated events start. Note that as found in simulations \cite{denny} this temperature is 
close to $T^{0}_{c}$. The value of $\tau_{o}$ is fixed in such a 
way that at $T=280K$, both the MCT and the hopping dynamics together predict a relaxation time 
which is close to that obtained in the experiments \cite{richert}. Thus for Salol system 
we find $\tau_{0}=2400 ps$.
The barrier height can also be calculated by considering the shape of the nucleating region 
to be a fuzzy sphere \cite{stevwoly}. The distribution of barrier height arises due to the fluctuation in entropy which can be related to the specific heat according to the Landau formula, $<(\Delta S)^{2}>=k_{B}C_{p}$,\cite{landau} where $\Delta S$ is the entropy fluctuation and $C_{P}$ is the specific heat. This expression can be rewritten in terms of 
configurational entropy fluctuation per bead $\delta s_{c}$ and heat capacity jump per bead ${\bar \Delta}c_{p}(T)$ as, $\delta s_{c}=\sqrt {\frac {k_{B} {\bar \Delta}c_{p}(T)}{\frac{4\pi}{3} (r^{\star}/a)^{3}}}$ and  $\frac{\delta s_{c}}{<s_{c}>}=\sqrt{\frac{3k_{B}TT_{K}}{4\pi s_{fit} (T-T_{K})^{2} (r^{\dagger}/a)^{3}}}$. Here $r^{\star}$ is the droplet radius determined using $F(r^{\star})=0$. 
$a$ is the length of the bead. We can also relate the entropy fluctuation to the width of the barrier height distribution, $\frac{\delta \Delta F}{\Delta F_{o}} \simeq \frac{\frac{\delta s_{c}} {<s_{c}>}}{1+{\frac{\delta s_{c}}{<s_{c}>}}}$, and thus define the width of the static barrier height distribution in terms of entropy fluctuation and specific heat.

Note that from the main equations of motion (eq.\ref{fqtmctdist} and eq.\ref{fqtfull}) it is obvious  that the hopping dynamics interacts with the MCT dynamics and thus effects the MCT dynamical correlation. But, the effect of MCT dynamics on the hopping motion is not so obvious. However, we will show below 
that the interaction between $\phi_{MCT}$ and $\phi_{hop}$ generates a renormalized barrier height 
distribution that is explored by the system. A treatment consistent with this renormalized barrier height distribution 
has not been attempted, although results should not be too different from the ones presented here. 

\section{Numerical Results}

We solve  equations \ref{fqtmctdist} and \ref{fqtfull} numerically using a numerical method presented earlier \cite{hofac} with a minor modification \cite{sbp2}.
The essential idea involved in the method is to separate the slow and the fast 
variables and treat them differently in the convolution. 
The short time part of $\phi(t)$ and $\phi_{MCT}(t)$ are calculated exactly
with very small step-size \cite{nume} and they are then used as input to carryout 
the calculation for the long time part of the same. 
The total relaxation time, $\tau_{total}$ and the stretching parameter, 
$\beta_{total}$ is obtained from the coupled dynamics by fitting the long time part of 
$\phi(t)$ 
(obtained from eq.\ref{fqtfull}) to a KWW stretched exponential function, $\phi(t)=A exp(-(t/\tau_{total}))^{\beta_{total}}$ 

\begin{figure}
\includegraphics[width=9cm]{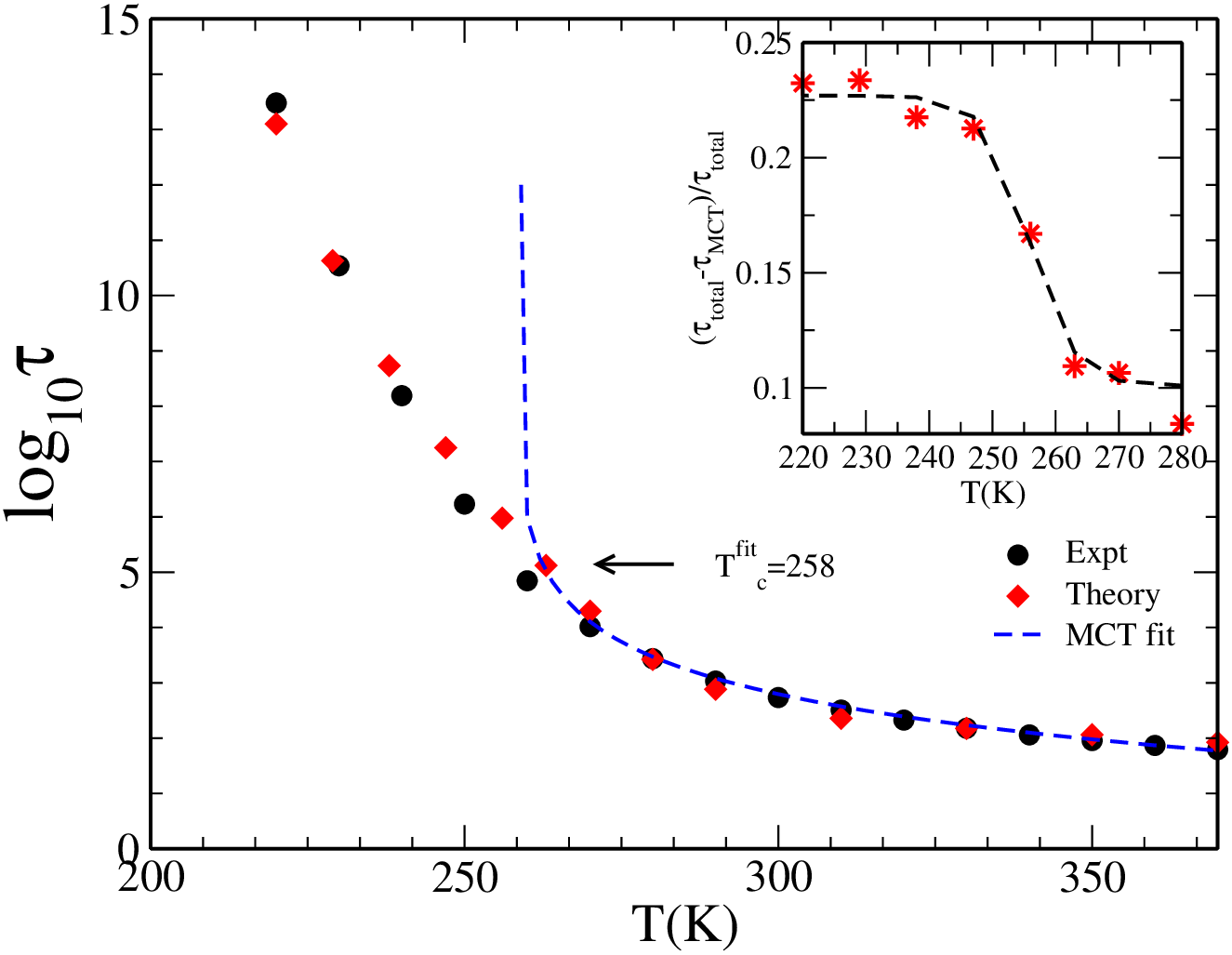}
\caption{Temperature dependence of the time scale of longtime $\alpha$ relaxation for Salol.
The timescale for the longtime part of the total structural relaxation, 
$\phi(t)$, obtained from experiments \cite{richert} (black circles) 
and that calculated from the present coupled theory, $\tau_{total}$ (red diamonds) 
are plotted against temperature. We fit the $\tau_{total}$ to the MCT phenomenological expression, $\tau_{total}\sim\left[(T-T^{fit}_{c})/T^{fit}_{c}\right]^{-\gamma}$ (blue dashed line), we obtain $T^{fit}_{c}=258K$. In inset we plot $(\tau_{MCT}-\tau_{total})/\tau_{total}$ as obtained from the theory (red stars). The black dashed line is a guide to the eye. The function shows a jump around $T=258 K$. We find that across this temperature the hopping dynamics changes its role and below $T^{fit}_{c}=258 K$ it plays a "direct" role in the dynamics}\label{tauT} 
\end{figure}

The plot for the relaxation time $\tau_{total}$ is given in \textbf{Fig.\ref{tauT}} where we have also shown the experimental results (for Salol) and the fit to the MCT phenomenological expression. As evident from the figure, the temperature dependence of $\tau_{total}$ reveals 
several interesting physics. The $\tau_{total}$ compares well with the experimental results \cite{richert} and predicts the correct glass transition temperature, $T_{g}=220K$. Recall that in our calculation the ergodic to non-ergodic transition takes place at $T^{0}_{c}=278 K$ where the activated 
dynamics also becomes significant. Thus we know that the idealized MCT prediction breaks down at $T=278 K$ and below this temperature, its validity as such is unclear. However when we fit the relaxation time to MCT expression ($\tau\sim\left[(T-T^{fit}_{c})/T^{fit}_{c}\right]^{-\gamma}$) we 
obtain a value, $T^{fit}_{c}=258K$, which is in excellent agreement with the experimental studies \cite{cu1,richert,fayer}. 

The advantage of the present scheme of calculation is that although it addresses the coupling between the continuous diffusion and the hopping motion, one can separately analyze their  relative contributions to the total dynamics and explore the origin of the apparent validity of the MCT below $T^{0}_{c}$. 
Due to the nonlinear coupling in Eqs.\ref{fqtmctdist} and \ref{fqtfull}, the activated dynamics plays both a "direct" 
and a "hidden" role in the total relaxation. The "direct" role is the direct relaxation of $\phi$ via $\phi_{hop}$. However, if we analyze the structure of Eq.\ref{fqtmctdist}, we find that the role of activated dynamics is also to soften the growth of the longitudinal viscosity which finally leads to the relaxation of the otherwise frozen $\phi_{MCT}$. Thus $\phi_{hop}$ plays a "hidden" role in the relaxation by helping $\phi_{MCT}$ to relax. We now analyze these two different roles of the hopping dynamics and their effect on the total relaxation. 

Our study shows that 
both continuous dynamics and activated dynamics changes continuously across $T^{fit}_{c}$. However, when we plot $(\tau_{MCT}-\tau_{total})/\tau_{total}$ against $T$ (in the inset of \textbf{Fig. \ref{tauT}}) we find that this quantity undergoes a jump like increase in the temperature range of the 
phenomenological MCT transition temperature $T^{fit}_{c}$. This implies that although below $T^{fit}_{c}$ the continuous diffusion 
still remains as the dominant mode of relaxation, there is a change of role and increased contribution of the activated dynamics. The plot suggest that in the range, $T^{fit}_{c}<T<T^{0}_{c}$, the activated dynamics plays only a "hidden" role whereas below $T^{fit}_{c}$ it also plays a "direct" role in the relaxation process.

Thus in the range, $T^{fit}_{c}<T<T^{0}_{c}$, the role of the activated dynamics is primarily to open up the bottle neck to start the diffusive dynamics. Once the latter starts, the relaxation primarily takes place via the diffusive channel and all the MCT predictions 
hold good. This provides a quantitative explanation of the observations reported by Kob {\it et al.} \cite{kob}. 
The authors find that till close to $T^{fit}_{c}$ the dynamics can be described via MCT although needs the static inputs
to be calculated at a higher effective temperature. 

Figure \ref{tauT} shows the important result that although idealized MCT
breaks down at $T^{0}_{c}=278K$, the functional form of MCT continues to describe 
the dynamical characteristics till $258K$. The latter is a widely known result, usually obtained from fitting the MCT functional forms to the experimental data \cite{cu1,richert,fayer}. Thus our study provides an explanation of this intriguing result in terms of the interaction between the MCT and activated dynamics which leads to \textit{hopping induced continuous diffusive motion.}
The study further predicts that below $T^{fit}_{c}$ although the continuous 
diffusive dynamics plays an important role in the relaxation, the increased contribution from the activated dynamics finally leads to the breakdown of the MCT predictions. 

\begin{figure}
\includegraphics[width=9cm]{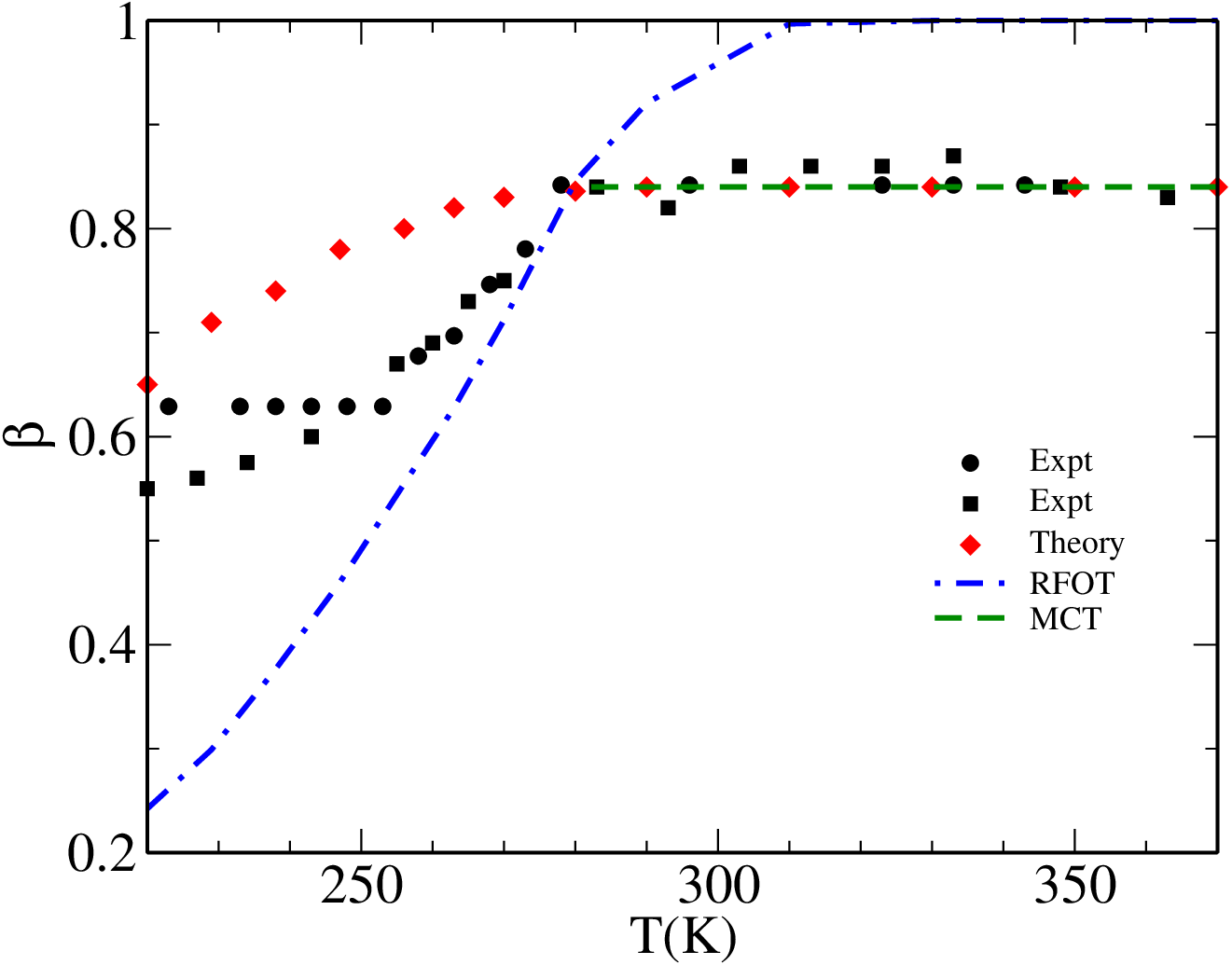}
\caption{Temperature dependence of the stretching parameter for Salol.
The stretching parameter for the longtime part of the total structural relaxation, 
$\phi(t)$, obtained from experiments \cite{cu1} (black squares) and \cite{cu2} (black circle)  
and that calculated from the present coupled theory, $\beta_{total}$, (red diamonds) 
are plotted against temperature. 
In the same plot we also present the stretching parameter predicted by the 
RFOT theory using the static barrier height distribution, (blue dashed-dot line) and the MCT (green dashed line) (which shows no temperature dependence of the stretching parameter).}\label{betaT}   
\end{figure}

In \textbf{Fig.\ref{betaT}} we plot the calculated temperature dependence of the stretching parameter $\beta$. In the same figure we also plot $\beta$ as predicted by the MCT and the RFOT theory (using static barrier height distribution) and also 
the experimental results on Salol \cite{cu1,cu2}.
It is clear that neither the MCT nor the RFOT can alone 
describe the proper temperature dependence of the stretching parameter. However,  
the coupled theory predicts a temperature dependence
of $\beta_{total}$ which is qualitatively similar to that found in experiments.
The stretching parameter is known to provide a measure of the heterogeneity in the system.
In our study the heterogeneity arises from the distribution of hopping barriers.  
However, below $T^{0}_{c}$, $\beta_{total}$ is smaller than $\beta_{RFOT}$ which means 
that the barrier height distribution which participates in the dynamics (dynamic barrier height distribution) is narrower than the initial assumption of distribution (static barrier height distribution). This modification of the barrier height distribution (or heterogeneity) takes place due to the coupling between the diffusive and activated dynamics.
The timescale analysis reveals that the smaller barriers contribute to the dynamics, which implies that the coupling leads to the facilitation effect. 
The proper temperature dependence of $\beta_{total}$ implies that the theory correctly predicts the modified barrier height distribution and thus the growth in dynamic heterogeneity.

\begin{figure}
\includegraphics[width=9cm]{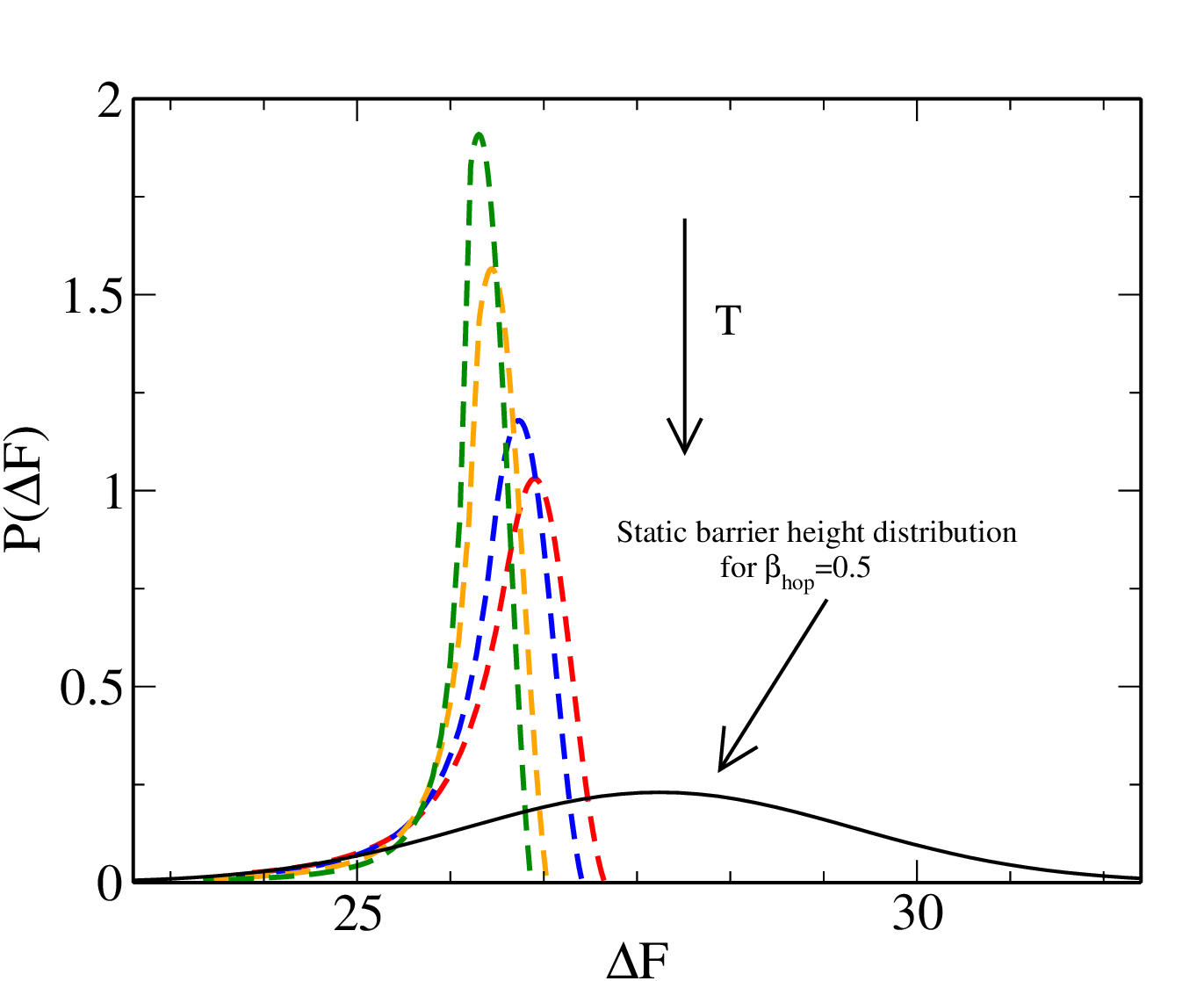}
\caption{Temperature dependence of the probability distribution of barrier heights which participate in the dynamics, ${\cal P}^{dynamics}(\Delta F)$. 
In this figure we plot the static barrier height distribution, ${\cal P}^{static} (\Delta F)$  which is used in the calculation of the total dynamics (black solid curve). From the total dynamics we again estimate the barrier height distribution 
which actually participates in the dynamics (colored dashed line). This has been done for different MCT relaxation time scale 
by changing the temperature of the MCT part of the dynamics (T=240K,230K,200K,175K). As the temperature is lowered the barrier height distribution becomes broader.}\label{barrier} 
\end{figure}

To quantify these points, we carry out the following analysis. Nonexponential relaxation in the system can be considered as arising from superposition of exponentials,
\begin{equation}
e^{-(t/\tau_{KWW})^{\beta_{KWW}}}=\int^{\infty}_{0} e^{-t/\tau} 
{\cal P}^{dynamic}(\tau) d\tau \label{dist} 
\end{equation}
\noindent
It is now possible to find the distribution function of the relaxation times
${\cal P}^{dynamic} (\tau)$ by Laplace inverting the stretched exponential.
Note that it is only possible to find ${\cal P}^{dynamic} (\tau)$ when the individual dynamics is exponential. 
Hence we do this analysis of the distribution of relaxation times (and thus the barrier heights which participate in the dynamics) considering only exponential MCT relaxation. 
Once we obtain the distribution of relaxation times we can convert it into the distribution of barrier heights using, ${\cal P}^{dynamic}(\Delta F) d\Delta F= {\cal P}^{dynamic}(\tau) d\tau$, provided $\tau$ can be
expressed in terms of $\Delta F$. The details are presented in the Appendix A.
We now analyze the dependence of ${\cal P}^{dynamic}(\Delta F)$ on the MCT relaxation time and show that ${\cal P}^{dynamic} (\Delta F)$ not only depends on ${\cal P}^{static}(\Delta F)$ but also depends on the MCT relaxation time
$\tau_{MCT}=K^{-1}_{MCT}$. We fix the ${\cal P}^{static}(\Delta F)$ and change the $\tau_{MCT}$ by calculating the MCT dynamics (which is coupled to ${\cal P}^{static}(\Delta F)$) at different temperatures. The results are presented in \textbf{Fig.\ref{barrier}}.

At high temperatures, where MCT dynamics is fast, ${\cal P}^{dynamic}(\Delta F)$ is narrow ($\beta_{KWW}$ is large), and the position of the maximum, $\Delta F_{max}$, is shifted to smaller $\Delta F$ (\textbf{Fig.\ref{barrier}}).  
We also find that as the MCT dynamics slows down, the distribution becomes broader and more non-Gaussian ($\beta_{KWW}$ also becomes smaller),
larger barriers contribute to the dynamics, 
and $\Delta F_{max}$ shifts to higher values (though it always remains smaller than $\Delta F_{o}$). 
These findings are in accord with the literature. We know that as the temperature is lowered (or the dynamics becomes slower) there is a growth in dynamic heterogeneity. Thus we may say that ${\cal P}^{dynamic}(\Delta F)$ provides a mean field estimation of the dynamic heterogeneity. 
Note that ${\cal P}^{dynamic}(\Delta F)$ is always narrower than ${\cal P}^{static}(\Delta F)$ 
and overlaps with it only in the low barrier side. Xia and Wolynes have given a physical interpretation of this exclusion of the higher barriers from the dynamics, using the picture of dynamic mosaic structure\cite{xiawoly}. Since the exclusion of higher barriers fastens the hopping dynamics thus this is termed as ``facilitation effect''. Our theoretical model does not explicitly consider these mosaic structures. However, we find that the presence of the continuous dynamics and its coupling to the activated dynamics leads to the facilitation effect. Note that the facilitation is strongest when the MCT dynamics is exponential and should be weaker when we have stretching in the MCT dynamics. 
\begin{figure}
\includegraphics[width=9cm]{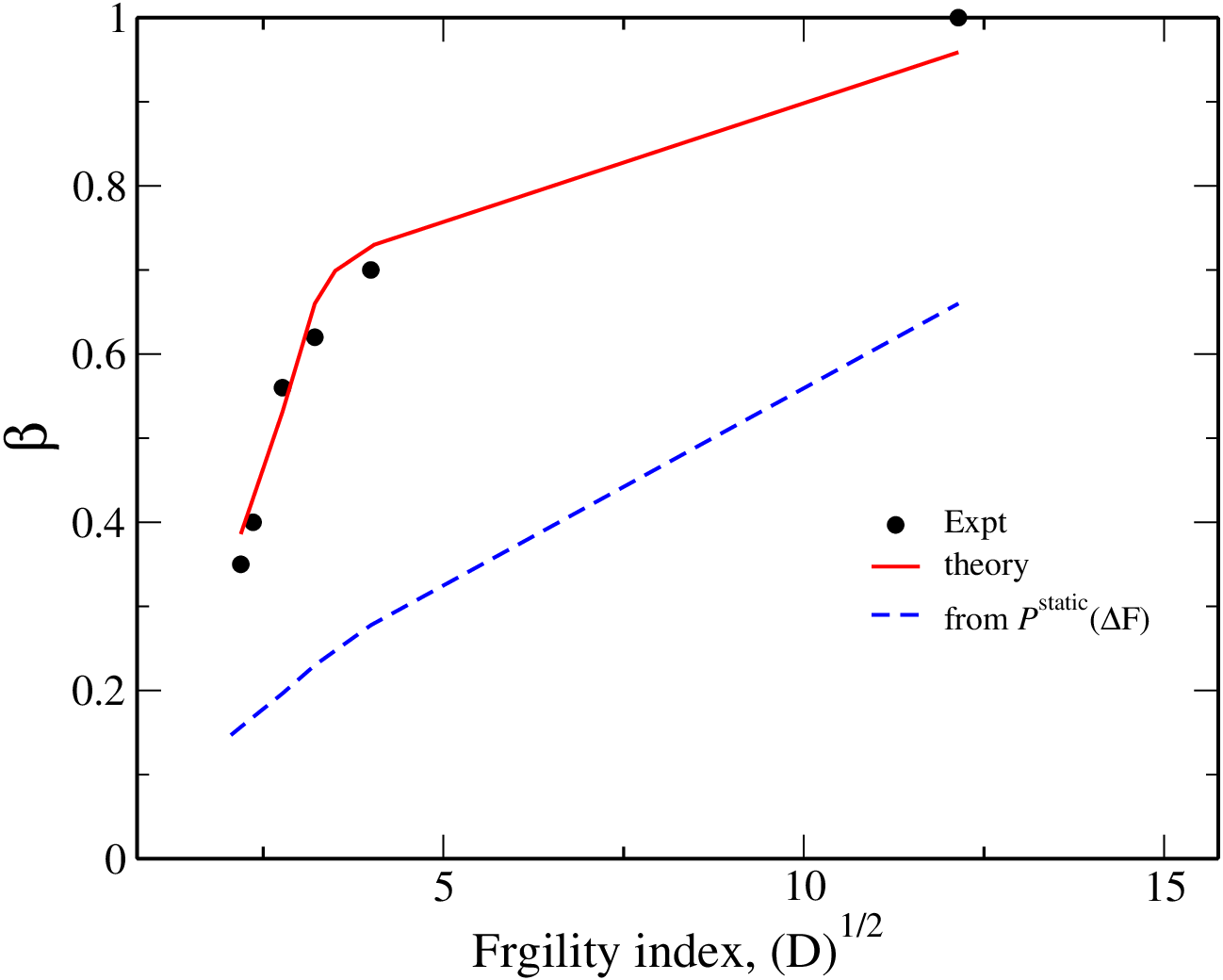}
\caption {Fragility dependence of the stretching parameter.
The stretching parameter for the total structural relaxation, 
$\phi(t)$, obtained from experiments $\beta_{expt}$ \cite{bohmer} (black circles) and that calculated from the present coupled theory, $\beta_{total}$ (red solid line) 
are plotted against the square root of fragility index $D$. As $D$ decreases the fragility of 
the system increases. In the same plot we also present the stretching parameter predicted by the full static barrier height distribution (blue dashed line).
The experimental and the theoretical values are at the glass transition temperature $T=T_{g}$. } \label{fragility}
\end{figure}

Next we show that this modification of the barrier heights, as predicted by the present theory plays a key role in describing the proper relationship between the fragility index, $D$, and the stretching parameter, $\beta$.
The study of Bohmer {\it et al.}\cite{bohmer} predicts a relationship between the fragility index $D$ and the stretching parameter $\beta$ at $T=T_{g}$. 
As discussed in Ref.\cite{xiawoly}, $\beta^{static}_{hop}$ is related to the width of the 
static Gaussian distribution of the barrier heights $\delta \Delta F$. $\delta\Delta F$ 
can be related to the fragility, D , $\frac{\delta\Delta F}{\Delta F_{o}}
\simeq\frac{1}{2\sqrt D}$. Thus $\beta^{static}_{hop}$ can be related to the fragility \cite{xiawoly}, $\beta^{static}_{hop}=[1+(\Delta F_{o}/2k_{B}T\sqrt D)^{2}]^{-1/2}$. However, it was found that this expression alone (with the 
full static barrier height distribution) does not describe the correct relationship between $\beta$ and $D$ \cite{xiawoly}. 

The details of the calculation are presented in Appendix B. In \textbf{Fig.\ref{fragility}} 
we plot the relationship between $\beta_{total}$ and $D$ as predicted by the present theory. 
In the same plot we also show the experimental results \cite{bohmer}. 
We find that the present theory captures the correct trend. 
This is to be contrasted with the predictions obtained from the static distribution of barrier heights. Note that in all the cases, $\beta_{total}>\beta^{static}_{hop}$, which implies that 
${\cal P}^{dynamic}(\Delta F)$ is always narrower than 
${\cal P}^{static}(\Delta F)$. Earlier we have shown that  
${\cal P}^{dynamic}(\Delta F)$ overlaps with ${\cal P}^{static}(\Delta F)$ only on the low barrier side. Thus the higher barriers do not participate in the dynamics leading to facilitation effect.
Hence, for a wide range of systems the theory predicts a proper modification of the barrier heights distribution. Note that it was shown by Xia and Wolynes that the static barriers height distribution needs a cut-off to describe the relationship between $\beta$ and $D$ \cite{xiawoly}. The authors gave a physical explanation of this "cutoff" using the picture of dynamic mosaic structure.

\section{Concluding Remarks}

In this article we present a theory that combines the effects of continuous dynamics (through mode 
coupling theory) and the activated dynamics (through random first order transition theory) 
in terms of a unified theory.
In this unified theory the two dynamics dose not just act as parallel channels of relaxation, but they interact with each other in a way that in the course of describing the dynamics they both modify each others behavior. Certain universal aspects of this work arise from this non-linear interactions. We show that the unified theory can provide satisfactory description of the relaxation over the whole temperature plane \cite{sbp} and
can explain the temperature dependence of the relaxation time, $\tau$ and the stretching parameter $\beta$. It also predicts the said relation between $\beta$ and D. 

The difficulty of performing MCT calculations for realistic systems is the non-availability of the static quantities. This difficulty can be overcome if we can estimate the ergodic to non-ergodic transition temperature $T^{0}_{c}$.
Following the analysis by Burmer and Reichman,\cite{reich} we assume that the ergodic to non-ergodic transition takes place at the temperature where the landscape energy first starts changing, $T_{L}$, thus $T^{0}_{c}=T_{L}$. We next use the observations reported by Sastry \textit{et al} that $T_{L}$ coincides with the emergence of non-exponential relaxation \cite{sastry}. From experimental studies we obtain  $T^{0}_{c}=T_{L}=278 K$ \cite{cu2}. In accord with simulation studies,\cite{denny}
in the present theory, the activated dynamics becomes significant, close to $T^{0}_{c}$. 
Thus in our theory, below $T^{0}_{c}$, the MCT can only relax with the aid of the
activated dynamics. In this theory, below $T^{0}_{c}$ the activated dynamics plays both a "hidden" and a "direct" role in the relaxation. The "direct" role is when it directly helps in structural relaxation and the 
"hidden" role is when it primarily lowers the longitudinal viscosity thus helping the otherwise 
frozen continuous diffusive dynamics to be active and help in the structural 
relaxation. The relaxation time $\tau_{total}$ obtained 
from the unified theory agrees well with experiments \cite{richert}. Although we consider the ergodic to non-ergodic transition and the activated dynamics to start at $T^{0}_{c}$, we find that when $\tau_{total}$ is fitted to MCT expression it shows a 
MCT transition temperature $T^{fit}_{c}=258 K$. We find that although the continuous and activated dynamics changes continuously across $T^{fit}_{c}$ the function $(\tau_{MCT}-\tau_{total})/\tau_{total}$ shows a jump across $T^{fit}_{c}$. This implies that hopping 
dynamics changes its role across $T^{fit}_{c}$.  In the range $T^{fit}_{c}<T<T^{0}_{c}$, it primarily plays a "hidden" role and the dynamics is well described by the MCT. This explains the study of Kob \textit{et al.} \cite{kob} and also the success of MCT in explaining experimental results below $T^{0}_{c}$ \cite{cummins1}. However, below $T^{fit}_{c}$, there is a sudden increase in the contribution from the activated dynamics and it plays a "direct" role in the relaxation, which results in the breakdown of the MCT predictions . \textit{Our study predicts that below $T^{fit}_{c}$, even though the MCT breaks down, the continuous dynamics persists and plays an important role in the relaxation}. Note that in an earlier simulation study we have shown the coexistence of the 
continuous and activated dynamics and we have further shown that activated dynamics, which is usually followed by continuous diffusive dynamics, relaxes 
the local stress \cite{sarikajcp}. Thus the simulations studies have already observed the "hidden" role of hopping which has been predicted in the present study.  

The theory can also predict the correct temperature dependence of the stretching parameter. Due to the growth of dynamic heterogeneity in the system the $\beta$ value is expected to decrease with $T$.
While MCT does not predict any temperature dependence of the stretching parameter and RFOT theory predicts a strong temperature 
dependence of $\beta^{static}_{hop}$, the prediction of the present theory is closer to the experimental results. The $T$ dependence of the $\beta_{total}$ arises due to the 
barrier height distribution of the activate dynamics. However, we find that the barrier height distribution which participates in the dynamics ${\cal P}^{dynamic}(\Delta F)$ gets modified from the static barrier height distribution, ${\cal P}^{static}(\Delta F)$ which is 
used as an input. The coupling of the MCT dynamics with the activated dynamics is found to 
lead to this modification. The higher barriers in ${\cal P}^{static}(\Delta F)$ are not present in ${\cal P}^{dynamic}(\Delta F)$, predicting a facilitation effect. We have shown that as the relaxation time of the system slows down, ${\cal P}^{dynamic}(\Delta F)$ becomes broader and includes higher barriers. Thus we predict that ${\cal P}^{dynamic}(\Delta F)$ provides a mean field estimation of the dynamic heterogeneity. 
We also find that this modification of the barrier height distribution, enables the theory to predict a proper relation between the fragility index $D$ and the stretching parameter, $\beta$.

{\bf ACKNOWLEDGEMENT}

This work was supported in parts from DST (J. C. Bose) and NSF (USA). SMB thanks Prof. K. Miyazaki for discussions.

\begin{large}\textbf{Appendix A}\end{large}

According to the present theory, $\tau=\frac{1}{K_{MCT}+K_{hop}}$
where $K_{hop}=\frac{1}{\tau_{0}}exp(-\Delta F/k_{B}T)$. Even for the simple exponential MCT relaxation
the relationship between $K_{MCT}$ and $\Delta F$ is non trivial \cite{sbp2}. However, in the low temperature limit we can write $K_{MCT}=\frac{2K_{hop}}{\lambda_{2}A^{2}+2\lambda_{1}A-1}=\alpha K_{hop}$, 
where $\alpha=\frac{2}{\lambda_{2}A^{2}+2\lambda_{1}A-1}$. $A$ is the Debye-Waller factor or the form factor (height of the plateau), thus, $\tau=\frac{1}{(\alpha +1)K_{hop}}$.
As discussed before,  ${\cal P}^{static}(\Delta F)$ is Gaussian. The mean barrier height, $\Delta F_{o}$, is calculated at $T=T_{g}$, using Salol parameters. The width of the distribution, $\delta \Delta F$ is related to the stretching 
parameter, ($\beta^{static}_{hop}=[1+(\delta\Delta F/k_{B}T)^{2}]^{-1/2}$ \cite{xiawoly,xiawoly28}) and it is taken in such a way that it gives $\beta^{static}_{hop}=0.5$. With these above mentioned parameters we 
calculate the total relaxation and fit it to a stretched exponential. From the fitting we obtain $\tau_{KWW}$ and $\beta_{KWW}$. This is now fed into eq.\ref{dist} to obtain ${\cal P}^{dynamic}(\Delta F)$.

\begin{large}\textbf{Appendix B}\end{large}

Since it is found that the mean barrier height is dependent on the configurational entropy and almost independent of the system thus for all the systems \cite{stevwoly}, we 
calculate $\Delta F_{o}$
using Salol parameters ($s_{fit}=2.65, T_{K}=175 K$, $T^{0}_{c}=278K$). The width of the barrier height distribution is fixed according to the fragility index $D$ which is taken from Bohmer et al. \cite{bohmer}. This then lets us define for each system, the static barrier height distribution and also the stretching in the hopping dynamics. For the Salol system we found that $\tau_{o}$ has to be used as a fitting  parameter to describe the proper timescale of the dynamics. In this calculation we also vary $\tau_{o}$ for each system in such a way that at $T=T_{g}$ the relaxation time is of the order of $100s$. 
We find that for more fragile system we need a larger value of $\tau_{o}$

We now need to define the corresponding MCT dynamics for each system. 
 Above the MCT transition temperature $T^{fit}_{c}$, 
the dynamics is expected to be described completely by the MCT dynamics. 
Thus the $\beta_{total}$ above $T^{0}_{c}{L}$ should be equal to the $\beta_{MCT}$ value. 
For the systems like Salol \cite{cu1}, Glycerol \cite{gly} and Silica \cite{silica} where the value of the stretching parameters at high temperatures are know we use those values to find the MCT parameters. For the systems where the stretching parameter at high temperatures are not known we use a reasonable value (definitely larger than $\beta_{expt}(T=T_{g})$ and also vary it in small amount along with $\tau_{o}$ to get the proper relaxation timescale at $T_{g}$. As for the value of $T^{0}_{c}$ and $T_{K}$, we keep them same as the Salol system. The $\beta_{expt}$ values are again obtained from Bohmer et al. 
for the corresponding $D$ values which are used to describe the hopping dynamics.
So our only constrain is that the timescale should be of the order of $100s$ at $T_{g}$.


\begin{references}
\bibitem{correc} In the previous paper \cite{sbp} there was a mistake in the derivation of the hopping kernal. The correct expression is presented here. The details of the derivation will be presented elsewhere.
\bibitem{nume} The results converge properly for the total timestep N=510 and the initial timestep for the integration is taken to be $10^{-9} ps$.
\bibitem{silica} Silica is known to be a strong liquid with $\beta \simeq 1$.
\end{references}
\end{document}